\documentclass[conference]{IEEEtran}
\IEEEoverridecommandlockouts

\usepackage{cite}
\usepackage{booktabs}
\usepackage{array}
\usepackage{multirow}
\usepackage{longtable}
\usepackage{graphicx}
\usepackage{hyperref}
\usepackage{enumitem}
\usepackage{amsmath}
\usepackage[table]{xcolor}
\usepackage[font=footnotesize,labelfont=bf]{caption}

\IfFileExists{tcolorbox.sty}{\usepackage[most]{tcolorbox}}{}
\IfFileExists{fontawesome5.sty}{\usepackage{fontawesome5}}{%
\newcommand{\faFileAlt}{\textbf{[D]}}
\newcommand{\faTimesCircle}{\textbf{[X]}}
\newcommand{\faCheckCircle}{\textbf{[OK]}}
}
\providecommand{\faFileAlt}{\textbf{[D]}}
\providecommand{\faTimesCircle}{\textbf{[X]}}
\providecommand{\faCheckCircle}{\textbf{[OK]}}

\IfFileExists{tcolorbox.sty}{%
\newtcolorbox{AuditBox}[2][]{
    enhanced,
    title={##2},
    colback=white,
    colframe=gray!60,
    coltitle=white,
    fonttitle=\bfseries\footnotesize,
    boxrule=0.5pt,
    arc=2pt,
    left=4pt,
    right=4pt,
    top=4pt,
    bottom=4pt,
    width=\linewidth,
    nobeforeafter,
    ##1
}
}{%
\newenvironment{AuditBox}[2][]{%
\setlength{\fboxsep}{4pt}%
\fbox\bgroup\begin{minipage}[t]{0.96\linewidth}\raggedright\footnotesize\textbf{##2}\par\smallskip
}{%
\end{minipage}\egroup
}
}

\title{MAG-Bot: A Multi-Agent Auditing Framework for Social Bot Detection}
\author{\IEEEauthorblockN{Sichen Zhao\textsuperscript{*}}
\IEEEauthorblockA{\textit{College of Engineering}\\
\textit{Northeastern University}\\
Boston, USA\\
zhao.siche@northeastern.edu}
\and
\IEEEauthorblockN{Yalun Qi}
\IEEEauthorblockA{\textit{Khoury College of Computer Sciences}\\
\textit{Northeastern University}\\
Boston, USA\\
qi.yal@northeastern.edu}
}

\begin{document}

\maketitle

\begin{abstract}
This paper studies social bot detection as dossier-based account auditing with large language models and a graph-structured multi-agent framework. From TwiBot-22, we reconstruct graph data into account-level records combining profile metadata, behavioral statistics, contextual cues, and recent tweets. We compare conventional feature-based baselines, a direct zero-shot Single-LLM auditor, and MAG-Bot, a LangGraph-based multi-agent system.

Three findings emerge. First, zero-shot Single-LLM auditing is feasible but has recall-related blind spots, especially on sparse, weakly grounded accounts and coherent role-bound personas. Second, role-constrained multi-agent decomposition substantially improves over Single-LLM: on the 585-account test split, MAG-Bot improves accuracy from 0.5846 to 0.7017, recall from 0.5986 to 0.8289, and F1 from 0.6747 to 0.8028. Third, the gain comes mainly from diagnosis-driven strengthening of the behavioral and contextual specialists, not aggregation tricks or post-hoc debate. Multi-agent LLM auditing therefore derives its main value from role-constrained evidence decomposition and blind-spot correction.
\end{abstract}

\section{Introduction}

Social bot detection has traditionally used graph-based, feature-based, and supervised classification approaches. Because social bots endanger online ecosystems and broader information environments, reliable detection is both a technical and societal concern \cite{ferrara2016rise}. These methods can perform strongly, but usually compress heterogeneous account evidence into one label; practical auditing often requires an inspectable evidence trail. This distinction matters when analysts need to understand which evidence views support a decision, why an account was flagged, and where uncertainty remains.

Large language models allow a social media account to be treated as an auditable dossier rather than a feature vector or graph node. A direct LLM auditor can inspect profile metadata, posting behavior, contextual cues, and tweet content, but our experiments show that Single-LLM auditing often over-trusts coherent role-bound personas and under-reacts to weak but cumulative bot signals.

This paper frames social bot detection as \emph{dossier-based auditing} and asks whether role-constrained multi-agent decomposition improves over direct Single-LLM auditing. We reconstruct TwiBot-22 \cite{feng2022twibot22} into account-level audit records and compare conventional feature-based baselines, a direct zero-shot Single-LLM auditor, and MAG-Bot. On the 585-account test split, MAG-Bot improves mainly through recall recovery, and mechanism analyses point to specialist strengthening rather than aggregation or debate. The paper makes four contributions:
\begin{enumerate}[leftmargin=*,itemsep=0pt,topsep=1pt]
    \item framing social bot detection as dossier-based auditing and reconstructing TwiBot-22 for feature-based and LLM-based evaluation;
    \item establishing a direct Single-LLM auditing baseline and showing non-monotonic prompt refinement;
    \item proposing MAG-Bot, a graph-structured multi-agent auditor with role-constrained specialists;
    \item showing that gains mainly come from specialist decomposition, not judge replacement or post-hoc debate.
\end{enumerate}

\section{Related Work}

This paper intersects supervised social bot detection, LLM-based account auditing, and multi-agent LLM reasoning. The key distinction from prior supervised detection work is that the present task emphasizes auditable evidence use rather than only final label prediction.

\subsection{Supervised Bot Detection}

Supervised bot detection treats the task as classification over metadata, content statistics, network structure, or their combinations. TwiBot-20 expanded public Twitter bot benchmarks and showed that many strong methods lose performance under more realistic conditions \cite{feng2021twibot20}; BotRGCN showed that heterogeneous relational graph modeling captures community structure and disguised bot behavior \cite{feng2021botrgcn}. TwiBot-22 extends this trajectory by addressing dataset scale, graph completeness, and annotation quality while supporting graph-based detection with diversified entities and relations \cite{feng2022twibot22}. These datasets and graph models provide necessary supervised reference points because they show what can be learned from structured metadata, behavior, and relational signals under fixed splits. The limitation for this paper is auditability: strong supervised models usually compress profile grounding, behavioral evidence, linguistic evidence, and contextual cues into one label, whereas an auditing setting requires those evidence views to remain inspectable.

\subsection{LLM-Based Auditing}

LLMs recast account assessment as inference over heterogeneous evidence and can return labels, rationales, evidence, and uncertainty notes. Feng et al. show that LLM-based bot detection can outperform strong baselines when adapted to heterogeneous user information, but LLM-guided manipulation can reduce detector reliability \cite{feng2024doesbotsayopportunities}. Direct LLM auditing is therefore promising but brittle: one prompt must aggregate multiple evidence views and may over-weight the aspect that first appears coherent or human-like. In this setting, fluent rationales can coexist with unstable decision boundaries, especially when accounts are sparse, weakly grounded, or role-bound. This concern matches broader evidence that fluent LLM reasoning can remain fragile, as in BRAINTEASER \cite{jiang2023brainteaser}.

\subsection{Structured and Multi-Agent Reasoning}

Structured LLM workflows separate evidence organization, intermediate analysis, and final judgment. DRP uses skill-aware step decomposition for reasoning-trace pruning \cite{jiang2025drp}, SCRIBE uses structured mid-level supervision to stabilize reasoning and tool use \cite{jiang2026scribestructuredmidlevelsupervision}, hybrid document-routed retrieval reduces cross-document confusion through LLM routing \cite{cheng2026resolvingrobustnessprecisiontradeofffinancial}, and TA-Mem separates memory extraction from retrieval-tool selection \cite{yuan2026tamemtoolaugmentedautonomousmemory}. For social bot detection, however, multi-agent design can add cost without improving evidence quality if specialists are weakly differentiated or the judge re-solves the task monolithically. Existing multi-agent work often bundles specialization, aggregation, critique, and debate, making the source of any gain hard to isolate. This study asks whether gains come from upstream specialist decomposition or downstream aggregation and debate.

\section{Dataset and Experimental Setup}

\subsection{Source Data and Reconstruction}

The experiments start from TwiBot-22 \cite{feng2022twibot22}, a large-scale graph-based Twitter bot detection dataset whose raw workspace includes user metadata, tweets, list information, hashtag information, split annotations, labels, and graph edges. Because the benchmark is organized around graph entities rather than account-level audit records, we reconstruct it into JSONL dossiers. Each record corresponds to one user and includes profile metadata, recent tweets, behavioral features, linguistic features, and contextual features. Profile fields include identity-facing attributes such as name, username, biography, location, verification status, and public metrics; derived summaries capture cadence, social footprint, repetition, lexical variety, and contextual grounding. The pipeline streams large JSON files, joins metadata and tweet shards, preserves original split identities, derives audit fields deterministically, and integrates edge-derived list information as compact contextual signal without requiring the full raw graph at inference.

\subsection{Audit Subset}

We constructed a balanced 4k working subset with 2{,}000 bots and 2{,}000 humans. Prompts cap recent tweets at 20 per account. Final LLM comparison uses only the preserved 585-account held-out test split; train and validation support development and reference analysis. The split is not re-balanced, so the test split is bot-heavy. The reconstructed subset contains 65{,}817 collected tweets in total and no missing profile records; its empty-bio and zero-tweet rates preserve the sparsity conditions that later matter for error analysis. Table~\ref{tab:dataset_stats} summarizes the split and quality checks.

\begin{table}[t]
\centering
\scriptsize
\begin{tabular}{lrrr}
\toprule
Statistic & Human & Bot & Total / Value \\
\midrule
Working subset & 2000 & 2000 & 4000 \\
Train & 1501 & 780 & 2281 \\
Validation & 335 & 799 & 1134 \\
Test & 164 & 421 & 585 \\
Tweet coverage & -- & -- & 88.73\% \\
Avg tweets / account & -- & -- & 16.45 \\
Avg tweets / nonzero account & -- & -- & 18.55 \\
Empty bio ratio & -- & -- & 23.18\% \\
Zero-tweet ratio & -- & -- & 11.28\% \\
\bottomrule
\end{tabular}
\caption{Dataset statistics and split composition for the balanced 4k account-level working subset. Final evaluation uses the 585-account test split.}
\label{tab:dataset_stats}
\end{table}

\subsection{Baselines and Prompt Setup}

All experiments are binary account classification: predict \texttt{bot} or \texttt{human}. Baselines are majority, logistic regression, and XGBoost. These feature-based models test whether the reconstructed account features contain predictive signal and provide a supervised reference point for the LLM systems. Because the preserved training split is human-dominant, the majority baseline becomes all-human and yields zero bot precision, recall, and F1. Single-LLM receives the full dossier and outputs structured JSON with decision, confidence, evidence, rationale, and uncertainty. Its prompt includes an explicit decision policy and fixed schema, asks the model to jointly inspect profile grounding, behavior, language, context, and tweets, and cautions against simplistic rules such as treating sparsity as bot evidence or persona coherence as human evidence. Table~\ref{tab:single_llm_development} shows why prompt-fixed \texttt{v2} is the Single-LLM reference and why prompt refinement is not monotonic. Version \texttt{v2} makes the direct auditor more willing to recognize accumulated weak bot signals, whereas broader boundary expansion in later variants hurts performance. Metrics are accuracy, precision, recall, F1, confusion matrices, and, for LLM systems, latency and token usage.

\begin{table}[t]
\centering
\scriptsize
\begin{tabular}{llccccc}
\toprule
Stage & Setting & Count & Acc. & Prec. & Rec. & F1 \\
\midrule
\multirow{4}{*}{Ablation} & Initial & 20 & 0.35 & 0.000 & 0.000 & 0.000 \\
& v2 & 20 & 0.60 & 0.700 & 0.583 & 0.636 \\
& v3 & 20 & 0.45 & 0.556 & 0.417 & 0.476 \\
& v3.1 & 20 & 0.55 & 0.636 & 0.583 & 0.609 \\
\midrule
\multirow{3}{*}{v2 stability} & Pilot & 20 & 0.60 & 0.700 & 0.583 & 0.636 \\
& Pilot & 50 & 0.56 & 0.750 & 0.583 & 0.656 \\
& Pilot & 100 & 0.56 & 0.764 & 0.575 & 0.656 \\
\bottomrule
\end{tabular}
\caption{Single-LLM prompt ablation and stability checks for \texttt{v2}.}
\label{tab:single_llm_development}
\end{table}

\section{MAG-Bot Methodology}

\subsection{Dossier Representation}

MAG-Bot operates on an account-level dossier, not a raw graph neighborhood or single text field. For user $u$, the shared Single-LLM/MAG-Bot input is
\[
x_u = \{p_u, b_u, l_u, c_u, t_u\},
\]
where $p_u$ is profile metadata, $b_u$ behavioral summaries, $l_u$ linguistic summaries, $c_u$ contextual cues, and $t_u$ capped recent tweets. This shared dossier representation supports a fair comparison between the monolithic Single-LLM path and the decomposed MAG-Bot path.

\subsection{Workflow and Specialist Reports}

\begin{figure}[t]
\centering
\IfFileExists{magworkflowfigure.jpg}{%
    \includegraphics[width=0.72\linewidth]{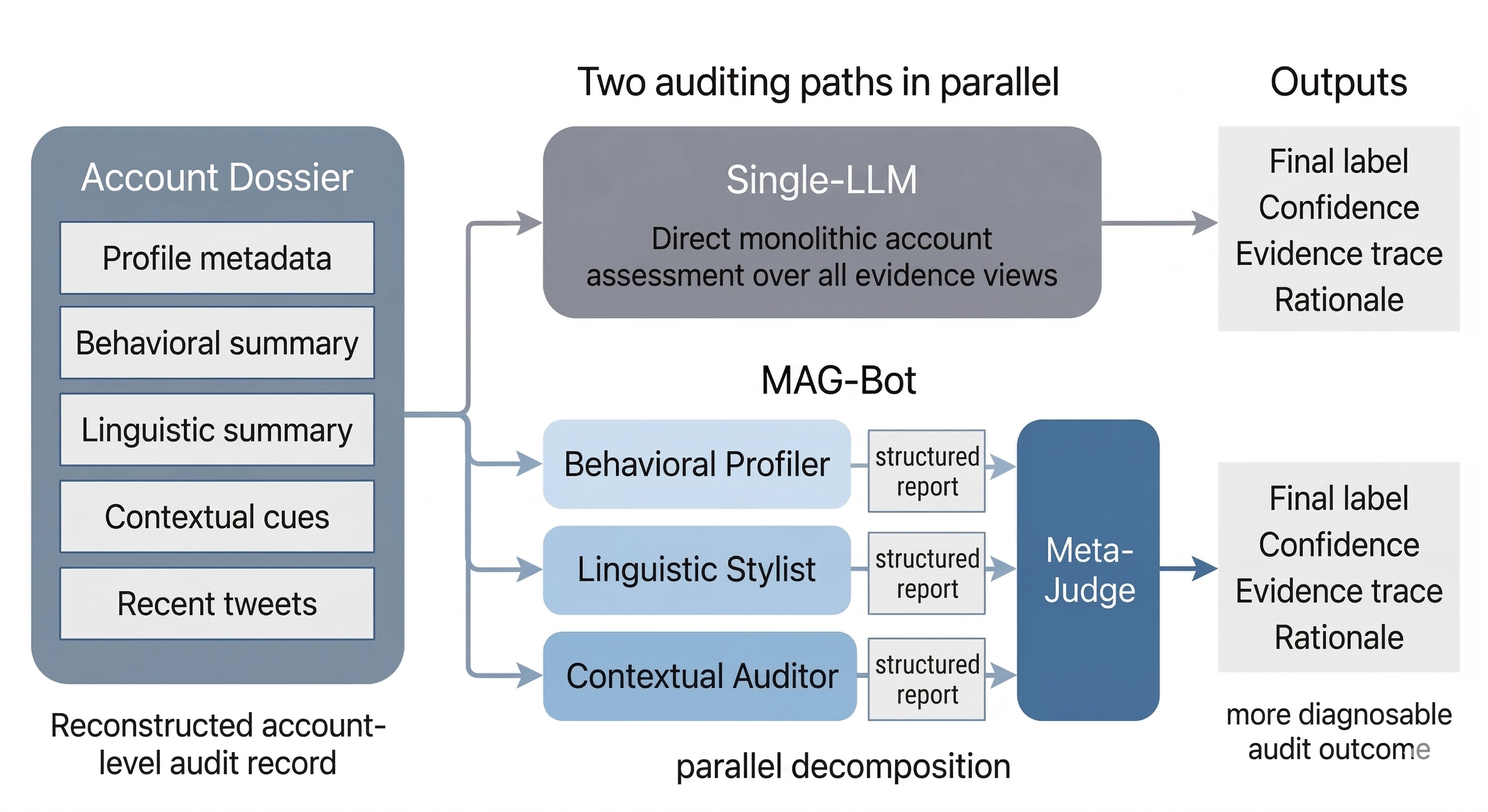}%
}{%
    \fbox{\parbox{0.88\linewidth}{\centering MAG-Bot workflow placeholder: dossier input, Single-LLM path, parallel specialists, Meta-Judge, and debate ablation branch.}}%
}
\caption{MAG-Bot workflow. The dossier is routed either to a direct Single-LLM auditor or to three specialists, each receiving role-specific evidence subsets, followed by a Meta-Judge.}
\label{fig:magbot_workflow}
\end{figure}

MAG-Bot is motivated by the fact that bot-detection errors arise from different evidence views, while a direct auditor resolves all evidence in one prompt and may overweight whichever view first appears human-like. Figure~\ref{fig:magbot_workflow} summarizes the workflow. MAG-Bot decomposes auditing into:
\begin{itemize}[leftmargin=*,itemsep=0pt,topsep=1pt]
    \item \textbf{Behavioral Profiler}: activity patterns, cadence, observability, and social footprint;
    \item \textbf{Linguistic Stylist}: wording variety, repetition, conversational tone, and broadcast style;
    \item \textbf{Contextual Auditor}: biography grounding, identity presentation, topical consistency, and contextual plausibility;
    \item \textbf{Meta-Judge}: final integration over specialist reports only.
\end{itemize}
Each specialist receives only role-relevant evidence: behavioral observability and activity statistics; recent tweets and linguistic summaries; or biography, identity fields, list cues, location, topics, and bounded recent content. These input boundaries reduce redundant global judgments and enable clearer diagnosis. They also make failures easier to localize: an error can be traced to behavioral permissiveness, linguistic over-trust, contextual over-skepticism, or judge integration. Each specialist emits
\[
o_u^{(k)} = \{y_u^{(k)}, c_u^{(k)}, e_u^{(k)}, r_u^{(k)}, z_u^{(k)}\},
\]
where $y$ is leaning, $c$ confidence, $e$ evidence, $r$ rationale, and $z$ uncertainty. The Meta-Judge receives only these structured reports, not the original dossier, so the final decision cannot collapse into a second monolithic audit. Operationally, specialists run in parallel, produce comparable reports, and leave final label integration to the judge. Debate is not part of the main method and appears only as an ablation.

\subsection{Specialist Strengthening}

MAG-Bot's main improvements came from diagnosis-driven strengthening of two specialists. \textbf{Behavioral v2} no longer treats the absence of obvious automation as positive human evidence and allows bot-leaning judgments when weaker signals accumulate, including sparse activity, weak social footprint, low observability, thin profile grounding, and limited interaction depth. \textbf{Contextual v2} no longer treats coherent persona presentation as sufficient evidence of a genuine human account and treats narrow, role-bound, academic, advocacy, and community-facing identities as potentially bot-leaning when lived personal grounding is weak. The linguistic specialist remains useful for style and repetition cues, but the main recall recovery comes from behavioral and contextual blind-spot correction. These revisions target the blind spots observed in staged diagnosis.

\section{Results and Analysis}

\subsection{Full-Test Results}

Table~\ref{tab:main_full_test} reports the full 585-account test comparison. Conventional supervised classifiers remain the strongest purely discriminative baselines, confirming that the reconstructed dossier features contain substantial predictive signal. Direct Single-LLM auditing is feasible but substantially weaker, especially in recall. MAG-Bot substantially improves over Single-LLM: accuracy rises by 11.71 points, recall by 23.03 points, and F1 by 12.81 points, with a slight precision gain. The pattern indicates recall recovery rather than a simple shift toward predicting more bots; the main methodological comparison is therefore between the two zero-shot LLM auditors.

\begin{table}[h]
\centering
\scriptsize
\setlength{\tabcolsep}{2pt}
\begin{tabular}{@{}p{0.34\columnwidth}cccc@{}}
\toprule
Method & Acc. & Prec. & Rec. & F1 \\
\midrule
Majority baseline & 0.2803 & 0.0000 & 0.0000 & 0.0000 \\
All-feature XGBoost & 0.7333 & 0.7422 & 0.9644 & 0.8388 \\
All-feature Log. Reg. & 0.7265 & 0.7277 & 0.9905 & 0.8390 \\
Single-LLM & 0.5846 & 0.7730 & 0.5986 & 0.6747 \\
MAG-Bot & 0.7017 & 0.7794 & 0.8289 & 0.8028 \\
\bottomrule
\end{tabular}
\caption{Final full-test results on the 585-account test split.}
\label{tab:main_full_test}
\end{table}

\subsection{Blind-Spot Recovery}

The gain should not be read as a generic effect of adding prompts. It reflects specialist exposure of evidence that a direct auditor underweights. The current best MAG-Bot corrects two persistent blind spots: over-trust in sparse or weakly grounded accounts, where the absence of strong human evidence was previously treated too leniently, and over-trust in coherent role-bound personas, where academic, professional, advocacy, or community-facing consistency was previously treated as overly strong human evidence. The trade-off is that MAG-Bot more readily challenges weak authenticity grounding, increasing false positives on some human-managed institutional accounts and narrow but genuine human presences. Thus the system improves bot sensitivity while exposing a boundary between automation-like managed presence and legitimate but tightly bounded human presence.

Figure~\ref{fig:case_study_audit} gives a representative held-out recovery case. It shows how MAG-Bot preserves cross-view conflict rather than letting linguistic plausibility mask behavioral and contextual thinness. The point is not only that the label changes, but that the audit trail makes the reason visible.

\begin{figure}[t]
\centering
\scriptsize
\begin{minipage}[t]{0.31\linewidth}
\begin{AuditBox}[equal height group=casecards,colframe=blue!50!black,colback=blue!4]{\faFileAlt~Dossier}
\textbf{@LTiroshi}; gold: \textcolor{red!70!black}{\textbf{BOT}}. Plausible name; Tel-Aviv; empty bio. Age $>$3.4 years; 8 profile tweets; 9 collected tweets; 44/98 followers/following. Tweets: ``Thank you Josh!'', ``Our paper is out!'' Surface: coherent academic persona.
\end{AuditBox}
\end{minipage}
\hfill
\begin{minipage}[t]{0.31\linewidth}
\begin{AuditBox}[equal height group=casecards,colframe=red!55!black,colback=red!4]{\faTimesCircle~Single-LLM}
\textbf{Decision: HUMAN} $(0.88)$; false negative. Plausible academic identity and location, natural interpersonal tone, and sparse activity read as low-volume human use. Failure: fluency and persona coherence masked thin grounding.
\end{AuditBox}
\end{minipage}
\hfill
\begin{minipage}[t]{0.31\linewidth}
\begin{AuditBox}[equal height group=casecards,colframe=green!45!black,colback=green!4]{\faCheckCircle~MAG-Bot}
\textbf{Decision: BOT} $(0.58)$; correct recovery. A: bot $(0.72)$, extremely thin footprint. B: human, polite authored language. C: mixed, coherent role but weak grounding. Meta-Judge: professional persona $\neq$ authentic personhood without grounding.
\end{AuditBox}
\end{minipage}
\caption{Qualitative audit case study. MAG-Bot recovers a bot missed by Single-LLM by preserving conflict between linguistic plausibility and weak behavioral/contextual grounding.}
\label{fig:case_study_audit}
\end{figure}

\subsection{Mechanism and Cost}

Earlier 100-sample diagnostics explain the full-test gain (Table~\ref{tab:mechanism_summary}). The broad-trigger debate activated on nearly all cases and acted as a human-protective smoothing stage, sharply reducing recall; stricter variants reduced this effect but still did not outperform MAG-Bot. This negative result suggests that preserving independent specialist evidence for final integration is more useful here than adding later-stage persuasion among agents. This aligns with Red Queen's caution that multi-turn structure can expose latent LLM vulnerabilities and should not be assumed beneficial \cite{jiang2025red}. In this task setting, decomposition helps, but post-hoc debate does not.

\begin{table}[t]
\centering
\scriptsize
\setlength{\tabcolsep}{2pt}
\begin{tabular}{@{}p{0.20\columnwidth}p{0.43\columnwidth}p{0.29\columnwidth}@{}}
\toprule
Analysis & Main finding & Implication \\
\midrule
Subset & Largest gains on sparse, weakly grounded bots and coherent role-bound bots; extra false positives on institutional or tightly bounded human-managed accounts. & Recall gain comes from targeted blind-spot recovery, not uniform score shifting. \\
Per-agent & Behavioral reports provide strongest bot-positive recovery; contextual reports prevent over-trust in plausible personas; linguistic reports support. & Specialist roles are differentiated. \\
Judge & Majority, confidence-weighted, and bot-sensitive rules underperform the LLM judge. & Simple aggregation does not recover the gain. \\
Debate & Debate variants increase interaction cost and remain below MAG-Bot. & Main MAG-Bot path remains the reference design. \\
\bottomrule
\end{tabular}
\caption{Mechanism analysis summary from earlier diagnostic studies.}
\label{tab:mechanism_summary}
\end{table}

MAG-Bot's quality gain is not free. On the full test split, Single-LLM requires 3.84 mean seconds per account and 1,281,758 total tokens, whereas MAG-Bot requires 7.98 seconds and 3,472,051 tokens, or approximately 2.08$\times$ latency and 2.71$\times$ token usage. MAG-Bot is therefore more bot-sensitive and diagnosable, but materially costlier. This trade-off should be interpreted as a different quality--cost operating point rather than a strict dominance claim over Single-LLM.

\section{Discussion and Conclusion}

\subsection{Main Claim}

This paper finds that a role-constrained zero-shot auditing workflow recovers many bots missed by a direct Single-LLM auditor, improving recall and F1 on the full held-out test split. The strongest improvements correct over-trust in sparse or weakly grounded accounts and coherent role-bound personas. Prompt refinement is not monotonic, judge replacement does not recover the same benefit, and debate does not improve MAG-Bot, suggesting that the main benefit comes from specialist decomposition and targeted blind-spot correction.

\subsection{Boundary and Future Work}

The framework nevertheless remains below the strongest supervised baselines. It should therefore be interpreted as an auditable LLM workflow with stronger recall than Single-LLM, not as the best possible classifier on a fully supervised benchmark. This is consistent with evidence that stronger neural or Transformer-based representations do not automatically dominate classical or simpler methods in every text-analysis setting: Lai et al. show that on short, noisy e-commerce reviews, classical embedding and topic-modeling methods can offer competitive or superior cost-performance trade-offs compared with Transformer-based sentence embeddings \cite{11484350}. LLM-based auditing should therefore be evaluated as a complementary, diagnosable workflow rather than a replacement for strong task-specific supervised baselines. Recovered bot recall also carries false-positive cost on some human-managed institutional accounts and weakly grounded genuine humans, while the quality gain entails higher latency and token usage. This trade-off matches prior concerns that automatic bot detection can produce unstable or inflated bot estimates across languages, thresholds, and evaluation settings \cite{rauchfleisch2020falsepositive}; bot-scoring outputs can support large-scale analysis but should not be treated as context-free ground truth without attention to uncertainty, intended use, and evaluation conditions \cite{yang2022botometer101}.

The contribution is not a universally best detector, but a more effective and diagnosable dossier-based LLM auditing framework that substantially improves over direct Single-LLM auditing while remaining interpretable at the evidence-view level. Future work should tighten false-positive control, improve specialist efficiency, broaden validation, incorporate graph signals through additional specialists, and test robustness against adversarial evidence manipulation. A future Graph Specialist would reconnect the dossier-based workflow more directly to the graph-rich structure originally provided by TwiBot-22. Although MAG-Bot is not pairwise rank aggregation, ranking-abuse work showing that MLE-based Bradley-Terry systems can be highly sensitive to structured perturbations motivates analysis of whether specialist reports, confidence scores, or judge-level aggregation rules are similarly vulnerable \cite{yao2026rankingabusestrategicpairwise}.

\bibliographystyle{plain}
{\small
\bibliography{references}
}

\end{document}